# Ferromagnetic quantum critical point in a locally noncentrosymmetric and nonsymmorphic Kondo metal

Soohyeon Shin[1,*,†], Aline Ramires[2,*,§], Vladimir Pomjakushin[3], Igor Plokhikh[1], and Ekaterina Pomjakushina[1]

[1] Laboratory for Multiscale Materials Experiments, Paul Scherrer Institut, Villigen 5232, Switzerland
[2] Condensed Matter Theory Group, Paul Scherrer Institut, Villigen 5232, Switzerland
[3] Laboratory for Neutron Scattering and Imaging, Paul Scherrer Institut, Villigen 5232 PSI, Switzerland

Corresponding authors: † soohyeon.shin@psi.ch, § aline.ramires@psi.ch
* These authors contributed equally to this work

**Abstract**

Quantum critical points (QCPs), zero-temperature phase transitions, are windows to fundamental quantum-mechanical phenomena associated with universal behaviour and can provide parallels to the physics of black holes. Magnetic QCPs have been extensively investigated in the vicinity of antiferromagnetic order. However, QCPs are rare in metallic ferromagnets due to the coupling of the order parameter to electronic soft modes [1,2]. Recently, antisymmetric spin-orbit coupling in noncentrosymmetric systems was suggested to protect ferromagnetic QCPs [3]. Nonetheless, multiple centrosymmetric materials host FM QCPs, suggesting a more general mechanism behind their protection. In this context, $CeSi_{2-\delta}$, a dense Kondo lattice crystallising in a centrosymmetric structure, exhibits ferromagnetic order when Si is replaced with Ag. We report that the Ag-substitution controls the strength of the Kondo coupling, leading to a transition between paramagnetic and ferromagnetic Kondo phases. Remarkably, a ferromagnetic QCP accompanied by concurrent strange-metal behaviour emerges. Herein, we suggest that, despite the centrosymmetric structure, spin-orbit coupling arising from the local noncentrosymmetric structure, in combination with nonsymmorphic symmetry, can protect ferromagnetic QCPs. Our findings present a unique example of Kondo coupling-driven ferromagnetic QCP through chemical doping and offer a general guideline for discovering new ferromagnetic QCPs.

## Introduction

Quantum critical points (QCP), zero-Kelvin phase transitions that are a consequence of the Heisenberg uncertainty principle, exhibit exotic quantum phenomena associated with order parameter fluctuations controlled by non-thermal tuning parameters [4]. Non-Fermi liquid (NFL) or strange metallic behaviour, a departure from the standard Fermi liquid (FL) behaviour associated with normal metals, generally appears near QCPs due to the strong scattering between conduction electrons and quantum fluctuations [5]. Strange-metal behaviour has been investigated in various antiferromagnetic (AFM) QCPs [6]. In contrast, QCPs in ferromagnetic (FM) compounds are rare, as they are often cut out by a first-order transition [7], or smeared out by inhomogeneity [8] or short-range order [9,10]. Theory developed by Belitz, Kirkpatrick, and Votja (BKV) showed that while in clean systems the FM quantum phase transition (QPT) is generally first order due to the coupling of the magnetization to soft electronic modes, in disordered systems the QPT can be associated with a second order phase transition, in other words, with a legitimate QCP [11,12]. Indeed, it was observed that the Curie temperature of the itinerant FMs $UCo_{1-x}Fe_xGe$ [13] and $Mn_{1-x}Cr_xSi_2$ [14] is driven to zero by doping, and the observed critical exponents corroborate the predictions of the BKV theory. Although rare, local-moment FM QCPs with concurrent NFL behaviour have been reported in clean systems such as $CeRh_6Ge_4$ and $YbNi_4(P_{1-x}As_x)_2$ [5,15,16].

More precisely, the avoided FM QCPs have been ascribed to soft modes or massless two-particle excitations of a generic Fermi liquid when spin-orbit coupling (SOC) is negligible [1,2]. The diverging spin susceptibility of the Fermi liquid at zero temperature and magnetic field contributes to the free energy with non-analytic terms in the magnetization, making the phase transition first order at low temperatures [1,2,17]. However, it has been suggested that strong spin-orbit coupling introduced by the breaking of spatial inversion symmetry in noncentrosymmetric materials introduces a cut-off to this divergence and can restore the second-order nature of the QPT [3]. Indeed, a FM QCP has been reported for $CeRh_6Ge_4$, a heavy fermion material that crystallizes in a noncentrosymmetric hexagonal structure ($P\bar{6}m2$, no. 187) [5]. However, a FM QCP has also been reported in $YbNi_4(P_{1-x}As_x)_2$, which adopts the centrosymmetric tetragonal structure ($P4_2/mnm$, no. 136) [15]. A better understanding of the key ingredients for the existence of FM QCPs is highly desirable.

$CeSi_{2-\delta}$ is the first dense Kondo FM in which ordered Ce moments are reduced due to Kondo screening. Figure 1a shows that Ce and Si atoms are located at 4a(0,3/4,1/8) and 8e(0, 1/4, 0.2910(2)) Wyckoff position, respectively, in the centrosymmetric $ThSi_2$-type tetragonal structure ($I4_1/amd$, no. 141) [18]. Stoichiometric $CeSi_2$ exhibits a paramagnetic ground state, with the Ce moments Kondo-screened by the conduction electrons, as schematically depicted on the left of Fig. 1b [19]. When Si-site deficiency ($\delta$) is larger than 0.15, the Ce moments order ferromagnetically below $T$ = 11 ~ 14 K, with the Curie temperature increasing with $\delta$ [20]. In addition, early pressure experiments on $CeSi_{1.81}$ suggested the existence of a FM QCP near P = 13.1 kbar [21]. However, the concurrent NFL behaviour associated with the $\delta$-driven putative FM QCP has not been investigated because of a structural transition [20]. Nonetheless, it has been reported that replacing Si with Ag also induces FM ordering without any structural transition [18]. Here, we report a potential

FM QCP and concurrent NFL behaviour by controlling Ag-doping in the dense Kondo lattice $CeSi_{2-\delta}$ and suggest the importance of local noncentrosymmetricity and nonsymmorphicity for the general stabilization of FM QCPs.

**Results**

Figure 1c is a magnetic phase diagram of $Ce(Si_{1-x}Ag_x)_{2-\delta}$ showing various magnetic ground states, *i.e.*, a paramagnetic (PM) state at $0 \leq x < 0.1$, a FM state at $0.1 \leq x < 0.22$, and an AFM state at $0.23 < x \leq 0.30$. The transition temperatures were determined by kinks or peaks of a derivative of magnetic susceptibility $\chi$, electrical resistivity $\rho$, specific heat capacity $C$, and imaginary part of AC susceptibility $\chi_{AC}$ (for details, see Extended Data Fig. E1). As shown in Fig. 1b, neutron powder diffraction experiments reveal that the Ce moments lie within a tetragonal plane in the FM state. On the other hand, ordered moments are perpendicular to the tetragonal plane in the AFM state (for details, see Extended Data Fig. E2, Table E1, and the Supplementary Information SI). Neutron diffraction experiments allow for an estimation of the magnitude of the ordered moments. For $x = 0.1$, the moments are estimated to be 0.71(2) $\mu_B$, smaller than the expected value of ~1.25 $\mu_B$ [22] expected for Ce ions in this crystalline environment. The reduced ordered moments are fully recovered in the antiferromagnetic state with 1.29(1) $\mu_B$ at $x = 0.35$, indicating that the Kondo hybridization (blue spheres in Fig. 1b) becomes weaker with increasing $x$. The FM order is suppressed to zero-temperature at a critical concentration $x_c \sim 0.07$, characterising a putative FM QCP [5,8,15].

Figure 1d shows the temperature-dependent magnetic susceptibility $\chi(T)$ of three different magnetic ground states. FM and AFM transition temperatures are indicated by $T_C$ and $T_N$, determined by a peak of $\partial\chi/\partial T$ and kink of $\chi(T)$, respectively. In the PM state, field-dependent magnetization $M(\mu_0 H$=7T) of $x$ = 0.1 is 0.42 $\mu_B$, indicating that the ordered Ce moments in the FM state are reduced to approximately 1/3 of $Ce^{3+}$ moment. As shown in Fig. 1e, the fully localised Ce moments in the AFM state ($x \geq 0.27$) are continuously suppressed with decreasing $x$ towards the putative FM QCP, indicating that the Kondo hybridization is systematically controlled by Ag-substitution. The continuous suppression of Ce moments and FM transition temperature with decreasing $x$ suggests that the FM order is controlled by the Kondo coupling strength. The size of the ordered moment (empty squares) deduced from the neutron experiments is in line with the tendency of $M(\mu_0 H$=7 T) (filled squares).

Figure 2a shows that the peak in $C/T$ due to the FM transition decreases with decreasing $x$, as indicated by arrows, reaching $T$ = 0 K near the critical concentration $x_c$ = 0.07. Bulk property measurements, $C/T$ and $\rho(T)$, do not show any feature of magnetic ordering down to 0.36 K at $x \leq x_c$, indicating the FM QCP at $x_c \sim 0.07$. As shown in Fig. 2b, large entropy accumulation is observed near $x_c$, with $C/T \sim 0.6$ J/mol/K$^2$ at $T$ = 0.36 K, comparable to other Ce-based heavy-fermion compounds [5,23,24]. $\Delta C/T$, obtained by subtracting $T^3$ contribution from $C/T$, follows a $-\log T$ dependence, the hallmark of NFL behaviour, for $x$ = 0.06 and 0.05, down to the lowest measured temperature. The NFL behaviour near $x_c$ is also supported by electrical transport data. Temperature-dependent electrical resistivity $\rho(T)$ of $x$ = 0 and 0.07 are plotted in Fig. 2c. For $x$ = 0, $\rho \propto T^2$ follows the standard FL theory, while for $x$ = 0.07, $\rho \propto T$, another hallmark of NFL behaviour in the temperature range of 0.45 K < $T$ < 5 K. Furthermore, the transverse magnetoresistance (MR) for $x$ = 0 at $T$ = 1 K exhibits an $H^2$ dependence, associated with normal-metal behaviour, whereas the MR for $x$ = 0.07 shows an $H$-linear dependence up to $\mu_0 H$ = 14 T. Similar behaviour has been reported for $FeSe_{1-x}S_x$ and $La_{2-x}Ce_xCuO_4$ due to quantum critical fluctuations [25,26]. The divergence of $C/T$, as well as $T$- and $H$-linear $\rho$ at $x \sim x_c$, indicates that the FM QCP appears in the dense Kondo ferromagnet $CeSi_{1.97}$ by Ag-substitution.

Figure 3a summarises the quantum critical behaviour around the FM QCP in $Ce(Si_{1-x}Ag_x)_{1.97}$. The upper panel of Fig. 3a shows that $C/T$ at $T$ = 0.34 K exhibits a peak at $x \sim x_c$. In addition, as plotted on the right ordinate in the upper panel of Fig. 3a, the exponent $n$ defining the temperature dependence of the resistivity, $\rho(T) = \rho_0 + AT^n$, is near unity, indicating a linear-in-temperature behaviour around $x_c$. The lower panel displays a colour plot of $n$, calculated from $\partial \ln(\rho-\rho_0)/\partial \ln T$, showing the expected funnel shape of NFL behaviour due to the magnetic quantum fluctuations, as reported in other FM and AFM QCPs [5,27]. The FL line in the phase diagram was obtained from the electrical resistivity measurements, below which $\rho(T)$ follows $T^2$ dependence (for details, see Extended Data Fig. E3).

## Discussion

Non-Fermi liquid behaviour can be observed in the absence of QCPs when metallic FMs are tuned by chemical doping [28-33] due to Griffiths phase effects [34]. Chemical doping induces inhomogeneity to the Kondo coupling strength, giving rise to Kondo-cluster-glass behaviour observed in $CePd_{1-x}Rh_x$ [32,33] and $CeCu_{1-x}Ni_x$ [31]. The scaling law of the quantum Griffiths phase is given by $\chi'_{AC} \propto C/T \propto T^{\varepsilon-1}$, $M \propto H^{\varepsilon}$, with $0 < \varepsilon < 1$ [29]. For instance, $CePd_{1-x}Rh_x$ exhibits the NFL behaviour without a diverging Grüneisen ratio, and experimental results are well explained by the above scaling laws [33]. However, $Ce(Si_{1-x}Ag_x)_{1.97}$ near $x_c$ exhibits a different scaling behaviour, with $\chi'_{AC} \propto T^{-1.7}$, $\Delta C/T \propto -\log T$, and $M \propto H^{0.5}$ (for details, see Extended Data Fig. E4). The absence of Griffiths phase is further confirmed by the absence of hysteresis between zero-field-cooled and field-cooled $\chi(T)$ in the temperature range where the $\rho(T)$ exhibits a $T$-linear dependence ($T < 5$ K).

Recent studies of Si-substitution in the FM quantum critical matter $CeRh_6Ge_4$ suggested that impurities might play an important role in the phenomenology of the FM QCP [35]. In contrast to the NFL behaviour near the pressure-induced FM QCP, the electrical resistivity near the critical Si concentration doesn't exhibit the $T$-linear behaviour although the Curie temperature seems to be suppressed continuously. At $x > x_c$, $CeRh_6(Ge_{1-x}Si_x)_4$ exhibits single-ion Kondo physics due to the inhomogeneous Kondo screening, which was not observed in $Ce(Si_{1-x}Ag_x)_{1.97}$. The $T$-linear $\rho(T)$ and the absence of inhomogeneous Kondo screening near $x \sim x_c$ suggest that the effects of the disorder are negligible in $Ce(Si_{1-x}Ag_x)_{1.97}$.

Based on the experimental evidence supporting the FM QCP in $Ce(Si_{1-x}Ag_x)_{1.97}$, we now address the origin of its stability. Recent work by Kirkpatrick and Belitz [3] showed how inversion symmetry breaking in (globally) noncentrosymmetric materials can provide a loophole to avoid the general mechanism that transforms a FM QCP into a first order phase transition. In noncentrosymmetric materials, antisymmetric spin-orbit coupling (SOC) splits the electronic bands in such a way that the two-particle excitations that couple to the magnetic order parameter acquire a mass and therefore do not contribute with nonanalytic terms to the free energy. This proposal lays out a natural explanation for the observation of signatures of FM quantum criticality in the noncentrosymetric heavy fermions $CeRh_6Ge_4$ [5,16] and UIr [36]. Nevertheless, previous work by the same authors suggests that this loophole is valid only for materials that have global inversion symmetry breaking [37-39]. If the material is only locally noncentrosymmetric, an effective chiral symmetry guarantees the double degeneracy of the electronic structure and therefore the presence of soft modes that can couple to the magnetization, leading to a first order phase transition in the same fashion as in Fermi liquids without SOC. One notable exception to this conclusion is the case of systems for which extra lattice symmetries guarantee that the coupling between the distinct chirality sectors is zero [37], which makes the material behave effectively as two independent globally noncentrosymmetric systems. In the presence of terms that couple the two chiral sectors, the nonanalyticity is restored, but its strength is reduced by a factor of $|\Delta(\boldsymbol{k})|^2/|\gamma(\boldsymbol{k})|^2$ [38,39] ($\Delta(\boldsymbol{k})$ corresponds to the coupling between the two chirality sectors and $\gamma(\boldsymbol{k})$ to the antisymmetric SOC). The calculations of Belitz and Kirkpatrick were done within the lowest order loop expansion [39]. Most recently, Miserev and collaborators [40] showed that second-order perturbation theory in the electron-electron interaction restores the nonanaliticities in the spin susceptibility, and therefore the first-order nature of the FM transition in noncentrosymmetric systems. Even though this result seems robust, the phenomenology presented here suggests that the prefactors accompanying such higher-order nonanaliticities are much smaller than for Fermi liquids in the absence of SOC, resulting in a weaker effect, allowing for the experimental observation of NFL behaviour around putative FM QCP in these systems.

Motivated by these observations, we constructed Table I. Remarkably, most of the heavy fermion FMs that display a second order phase transition and signatures of quantum criticality are either globally noncentrosymmetric, or locally noncentrosymmetric and nonsymmorphic. We, therefore, infer that nonsymmorphicity is a key ingredient in stabilizing these FM QCPs, as in nonsymmorphic materials the ratio $|\Delta(\boldsymbol{k})|^2/|\gamma(\boldsymbol{k})|^2$ can be made arbitrarily small if the Fermi surfaces are located close to the BZ boundaries. A rough trend of the ratio $|\Delta(\boldsymbol{k})|^2/|\gamma(\boldsymbol{k})|^2$ can be captured by the atomic distances between inequivalent sublattices sites. The larger the inter sublattice distances, the smaller the hopping amplitudes between the sublattices, what corresponds to a weaker coupling between sectors with opposite chirality. For the particular case of $CeSi_2$, Ce and Ce’ sublattices are not centers of inversion, despite the fact that the lattice is globally centrosymmetric. The center of inversion lies at the midpoints between Ce and Ce'. Furthermore, the two sublattices are related by a nonsymmorphic symmetry, as depicted in Fig. 3c (a more detailed discussion is given in the SI). The combination of these symmetries with a particular large distance between the inequivalent sublattices might be at the core of the protection of the QCP against its transmutation into a first order transition.

The trend presented here should trigger further theoretical work in order to validate its quantitative aspect. Conversely, the importance of noncentrosymmetricity and nonsymmorphicity can now also be used as general guidelines towards the protection of quantum critical behaviour in metallic systems.

## Tables

**Table I: Summary of known heavy-fermion FM displaying either a first or second order phase transition (QCP).** We highlight the reported type of QCP, in addition to the space group of each material, drawing special attention to the presence of absence of global (Glob.) and local (Loc.) inversion symmetry and nonsymmorphicity. In addition, the last column ($D_{sub}$) gives the distances between sublattices of lanthanide or actinide sites, which trend corroborates the discussion in the main text. Exceptions of the proposed picture are highlighted by (*) and (**). $YbNi_4(P_{1-x}As_x)_2$ and $U_4Ru_7Ge_6$, with the second order phase transition ascribed to the one-dimensional nature of the crystalline structure and a particularly large distance between inequivalent U sites relieving the need for nonsymmorphicity, respectively. Among the materials that display a first order phase transition, exceptions to this rule include U(Co, Rh)Ge and U(Co, Rh)Al, which can be understood in terms of the small distances between sublattices, what enhances the ISH. Another exception is $U_3P_4$, which is globally noncentrosymmetric and nonsymmorphic, but displays a first order phase transition. Despite the simple chemical formula, this system has a unit cell with exceptionally many atoms, which might make this simple picture not directly applicable.

| Material | Order | Type | Space group | Inversion Glob./Loc. | Symmorphic | $D_{sub}$ |
|---|---|---|---|---|---|---|
| $Yb(Cu,Ir)_2Si_2$ [41,42] | 1st | Local | $I4/mmm$ | Y/Y | Y | 3.924 |
| $UGe_2$ [43] | 1st | Itinerant | $Cmmm$ | Y/N | Y | 3.854 |
| U(Co,Rh)Ge* [44] | 1st | Itinerant | $Pnma$ | Y/N | N | 3.480 |
| U(Co,Rh)Al* [45,46] | 1st | Itinerant | $P\bar{6}2m$ | N | Y | 3.394 |
| $U_3P_4$ * [47] | 1st | Itinerant | $I\bar{4}3d$ | N | N | 3.841 |
| $Ce(Pd,Pt)_{1-x}Ni_x$ [48] | 2nd | Local | $Cmcm$ | Y/N | N | 3.868 |
| $CeSi_{1.81}$ [21] | 2nd | Local | $I4_1/amd$ | Y/N | N | 4.055 |
| $UNiSi_2$ [49] | 2nd | Local | $Cmcm$ | Y/N | N | 4.060 |
| $CeRh_6Ge_4$ [5,16] | 2nd | Local | $P\bar{6}m2$ | N | Y | 3.855 |
| UIr [36] | 2nd | Itinerant | $P2_1$ | N | N | 3.637 |
| $YbNi_4(P_{1-x}As_x)_2$** [15] | 2nd | Local | $P4_2/mmm$ | Y/Y | N | 3.857 |
| $U_4Ru_7Ge_6$ ** [50] | 2nd | Itinerant | $Im\bar{3}m$ | Y/N | Y | 4.144 |

**Figures and captions**

Figure 1

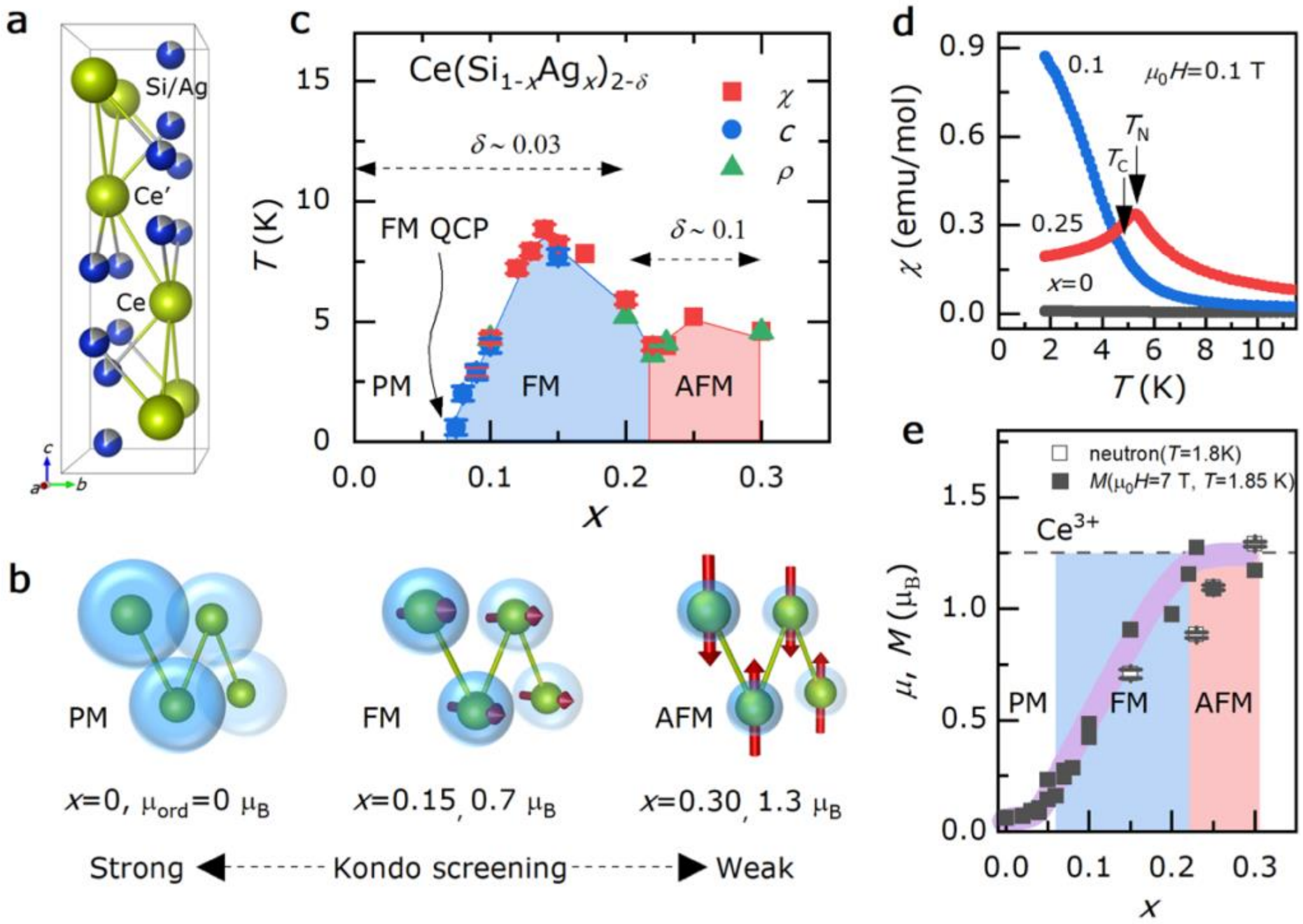

**Fig. 1 Magnetic phase diagram and Kondo coupling of $Ce(Si_{1-x}Ag_x)_{2-\delta}$. a,** Crystal structure of $Ce(Si_{1-x}Ag_x)_{2-\delta}$. Yellow, blue, and grey spheres represent Ce, Si, and Ag atoms, respectively. Ce and Ce' indicate two different Ce sublattices distinguished by different local atomic environments. **b,** Schematic drawings display magnetic structures of three different magnetic ground states, and the strength of Kondo screening (translucid blue spheres). $\mu_{ord}$ indicates the size of the ordered magnetic moments determined by neutron scattering experiments in units of Bohr magneton $\mu_B$. **c,** Magnetic phase transition temperatures, determined by measuring magnetic susceptibility $\chi$, specific heat capacity $C$, and electrical resistivity $\rho$, are plotted as a function of Ag-doping $x$. PM, FM, and AFM represent the paramagnetic, ferromagnetic, and antiferromagnetic states, respectively. **d,** Magnetic susceptibility for PM, FM, and AFM states is plotted as a function of temperature. $T_C$ and $T_N$ indicate transition temperatures of FM and AFM. **e,** The ordered magnetic moment (open squres) and magnetisation value (closed squares) at $\mu_0 H$ = 7 T and $T$ = 1.85 K are plotted as a function of $x$. The dashed line indicates the expected value for $Ce^{3+}$ of $Ce(Si_{1-x}Ag_x)_{1.9}$.

Figure 2

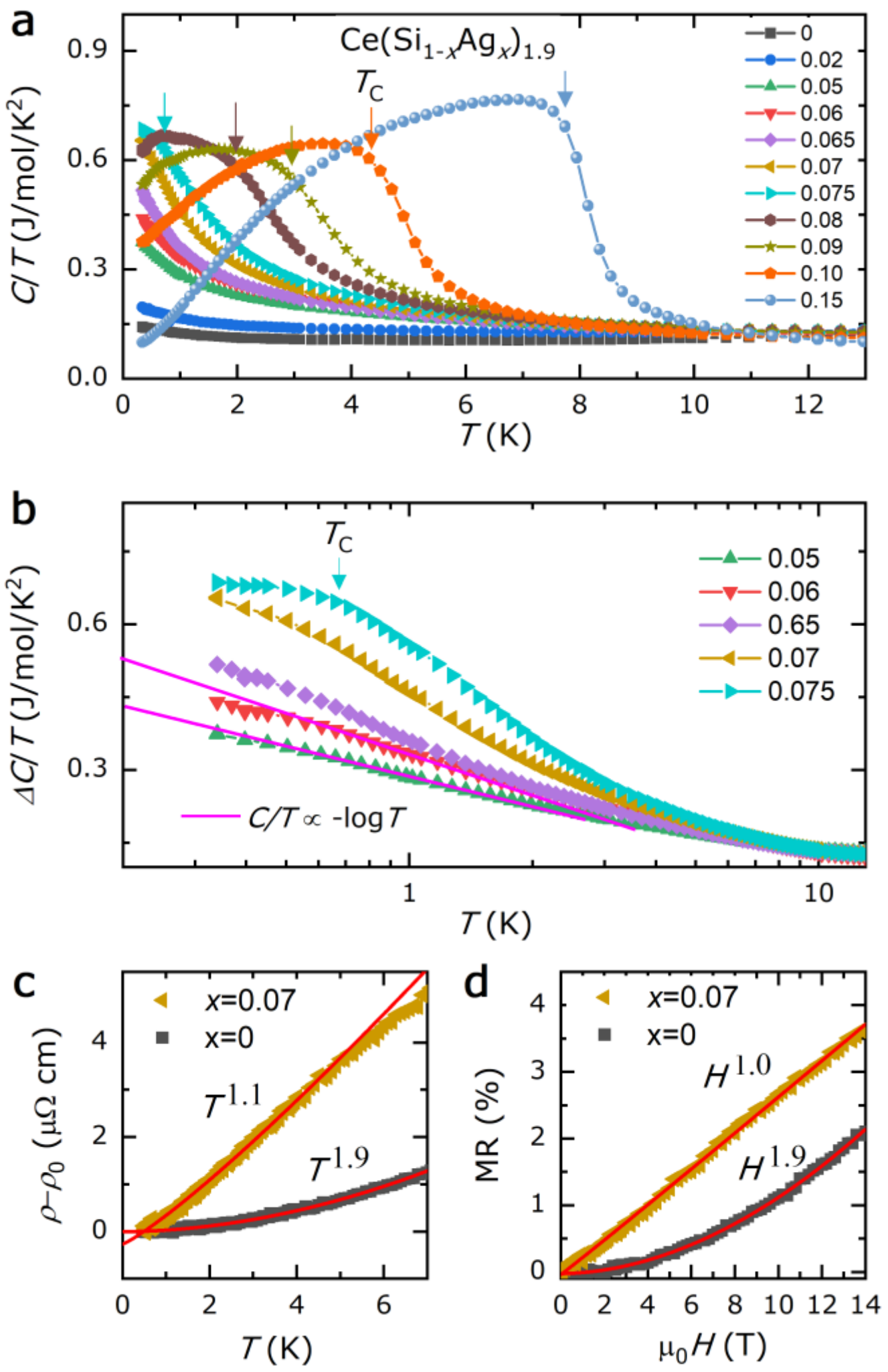

**Fig. 2 Non-Fermi-liquid behaviour near the FM QCP of $Ce(Si_{1-x}Ag_x)_{1.97}$. a,** Specific heat capacity divided by temperature *C/T* of $Ce(Si_{1-x}Ag_x)_{1.97}$ is plotted as a function of temperature. $T_C$ indicates the ferromagnetic transition temperature. **b,** $\Delta C/T$, obtained from subtracting $T^3$ term from *C*, is plotted as a function of temperature for *x* near the critical concentration. The magenta line represents $C/T \propto -\log T$ at low temperatures. **c,** Electrical resistivity $\rho$ after subtracting residual resistivity $\rho_0$ for $x = 0$ and 0.07 is plotted as a function of temperature. The red lines are the least-squares fitting using $\rho - \rho_0 \propto T^n$. **d**, Transverse magnetoresistance MR for $x = 0$ and 0.07 were measured up to $\mu_0 H = 14$ T at $T = 1$ K. The red lines are the least-squares fitting using MR $\propto H^{\beta}$.

Figure 3

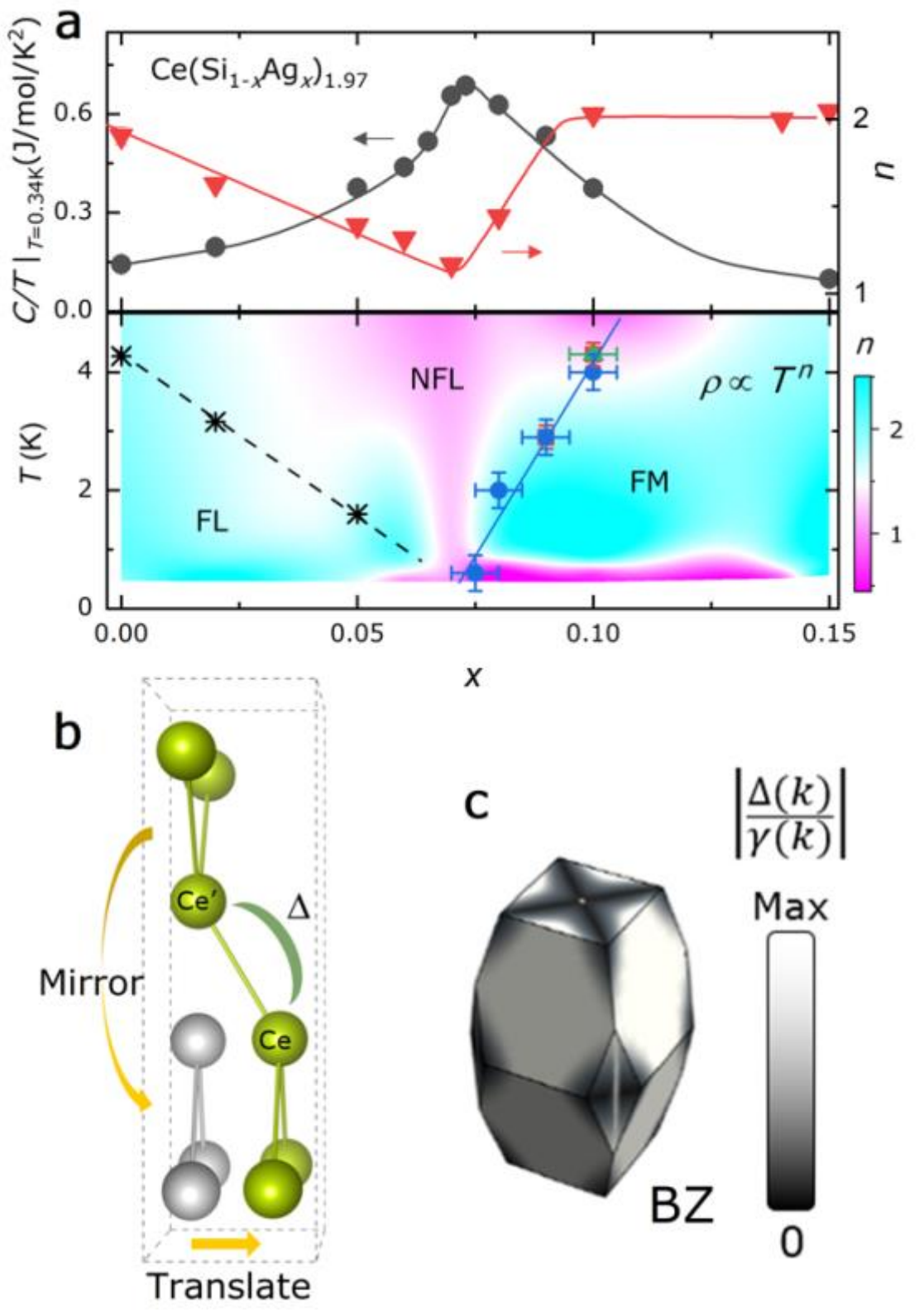

**Fig. 3 Quantum criticality near the FM QCP protected by nonsymmorphic symmetry. a,** In the upper panel, $C/T$ at $T$ = 0.34 K and resistivity temperature exponent $n$, estimated by the least-squares fitting using $\rho(T) = \rho_0 + AT^n$, are plotted as a function of $x$ on the left and right ordinate, respectively. In the lower panel, $T_C$ and $T_{FL}$ are plotted as a function of $x$ overlaid on the colour plot of $n$ on the $T - x$ plane. **b,** Crystal structure highlighting the two inequivalent Ce sublattices indicated by Ce and Ce', and the nonsymmorphic symmetry operation that takes one sublattice into another, a combination of mirror and translation symmetries. **c,** Ratio of inter-sublattice hopping ISH and spin-orbit coupling SOC plotted on the Brillouin zone (BZ) edges. Note that this ratio goes to zero along some lines (black colour) due to nonsymmorphic symmetries. The dark gray colour around these lines indicates regions within which the ISH dominates over SOC.

**Methods**

Materials synthesis and characterisation

Polycrystalline samples of $Ce(Si_{1-x}Ag_x)_{2-\delta}$ were synthesized by the arc-melting technique that was shown elsewhere [1]. Ce (rod, 99.9%, ChemPUR), Si (lump, 99.9999%, Alfa Aesar), and Ag (granule, 99.99%, ChemPUR) were prepared in a stoichiometric molar ratio. The weighted elements in stoichiometric composition were melted several times after flipping over to improve the sample homogeneity. To improve the crystal quality, the arc-melted buttons covered by Ta-foil were sealed in an evacuated silica tube for a thermal annealing at 850 °C for 10 days. Phase purity was investigated by powder x-ray diffraction measurements (PXRD) using a Bruker D8 Advance with Cu-cathode. All cases do not show any impurity phases. Crystal structure and stoichiometry were investigated by single-crystal diffraction using STOE STADIVARI diffractometer with Mo $K_\alpha$ radiation (0.71073 Å). For details of doping-dependent lattice parameters and chemical compositions, see Extended Data Table E2. All single-crystal x-ray diffraction results are displayed in SI. Note that the nominal $x$ was used for $0 \le x \le 0.20$ because the Si-site deficiency is negligible and lattice parameters are systematically controlled with nominal $x$, while actual $x$ was used for nominal value of $0.25 \le x \le 0.35$.

Neutron powder diffraction and refinement

The crystal and magnetic structures were studied by neutron powder diffraction using the high-resolution powder diffractometer for thermal neutrons (HRPT) [2] at the Swiss Spallation Neutron Source SINQ at Paul Scherrer Institut (PSI), Switzerland. About 2 g of each powder was loaded in 6 mm vanadium containers. Diffraction patterns were collected at temperatures of 1.8 and 15

K using neutrons with wavelengths of 1.494 and 2.45 Å. All diffraction data were analysed using the programs of the FullProf software suite [3]. The symmetry analysis of the magnetic structures was done using the Bilbao crystallographic server [4] and the ISODISTORT tool based on ISOTROPY software [5,6].

Bulk property measurements

Electrical resistivity measurements were performed using the standard four-probe (25 μm Pt wires) technique applying a current of 1 mA on the polished surface of bar-shaped specimens. Electrical resistance and heat capacity were measured by physical property measurement system (PPMS, Quantum Design) with He-3 insert. Magnetization measurements were performed on a superconducting quantum interference device (SQUID) installed in the magnetic property measurement system (MPMS, Quantum Design), in the temperature and magnetic field ranges from $T$ = 1.8 to 300 K and $\mu_0 H$ = 0 to 70 kOe, respectively.

## Acknowledgments

S.S. and E.P. would like to thank Eric D. Bauer and Hanoh Lee for the fruitful discussion. Neutron experiments were performed at SINQ in Paul Scherrer Institut, Switzerland. Electrical resistance, magnetization, and heat capacity measurements were performed in Laboratory for Multiscale Materials and Experiments at Paul Scherrer Institut. This project has been supported by the Swiss National Science Foundation SNSF Project No. 200021_188706. AR acknowledges support from the Swiss National Science Foundation through Ambizione Grant No. 186043.

## Author contribution

S.S., A.R., and E.P. initiated and lead the project. S.S. synthesised the crystals and performed the bulk property measurements. S.S. and V.P. performed the neutron experiments and analysed the results. I.P. performed and analysed the single-crystal x-ray diffraction experiments. S.S., A.R., and E.P. wrote the manuscript with input from all authors.

## Competing interest declaration

The authors declare no competing interests.

## Data availability

The data supporting this study are available via the Zenodo repository (DOI:10.5281/zenodo.8363352).

**Extended data figures and tables**

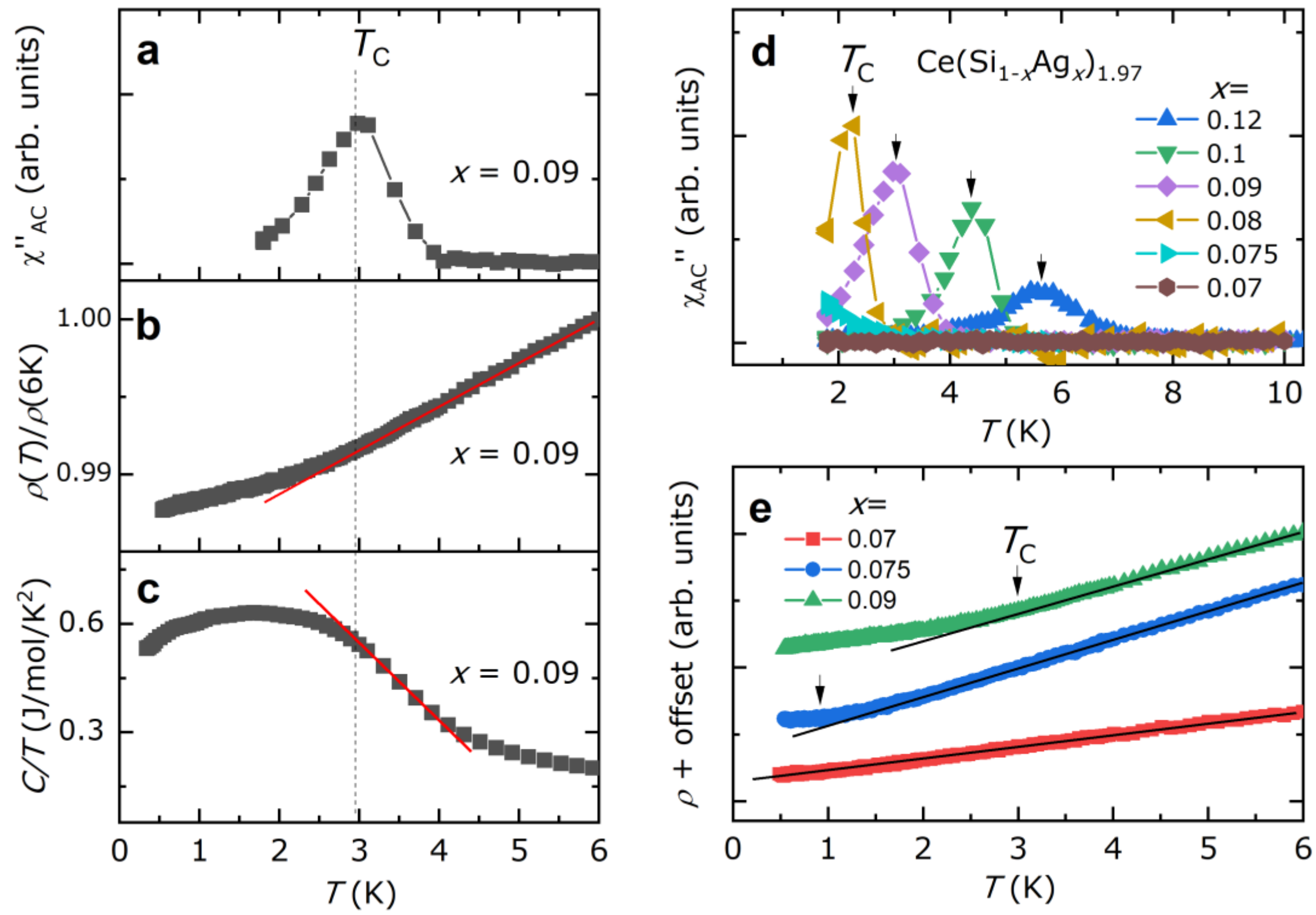

**Extended Data Fig. E1. Bulk property anomalies at the ferromagnetic phase transition. a,** Temperature-dependent imaginary part of AC magnetic susceptibility $\chi''(T)$ for $x = 0.09$ exhibits a peak at the ferromagnetic phase transition $T_C$. **b**, Temperature-dependent normalised electrical resistivity ρ($T$) changes a curvature below $T_C$. The red line indicates the linear behaviour above $T = T_C$. **c**, Temperature-dependent specific heat capacity divided by temperature $C(T)/T$ exhibits the FM transition below which $C(T)/T$ deviates from the steep increase, and this temperature is consistent with other FM anomalies. **d**, The FM transition peak in $\chi''(T)$ of various $Ce(Si_{1-x}Ag_x)_{1.97}$ shifts to lower temperatures with decreasing $x$. **e**, $\rho(T)$ of $x = 0.09$ and 0.075 shows the suppression of $T_C$ with decreasing $x$. $\rho(T)$ of $x = 0.07$ doesn't change curvature down to $T = 0.45$ K, showing the linear behaviour from the lowest measured temperature to $T = 6$ K.

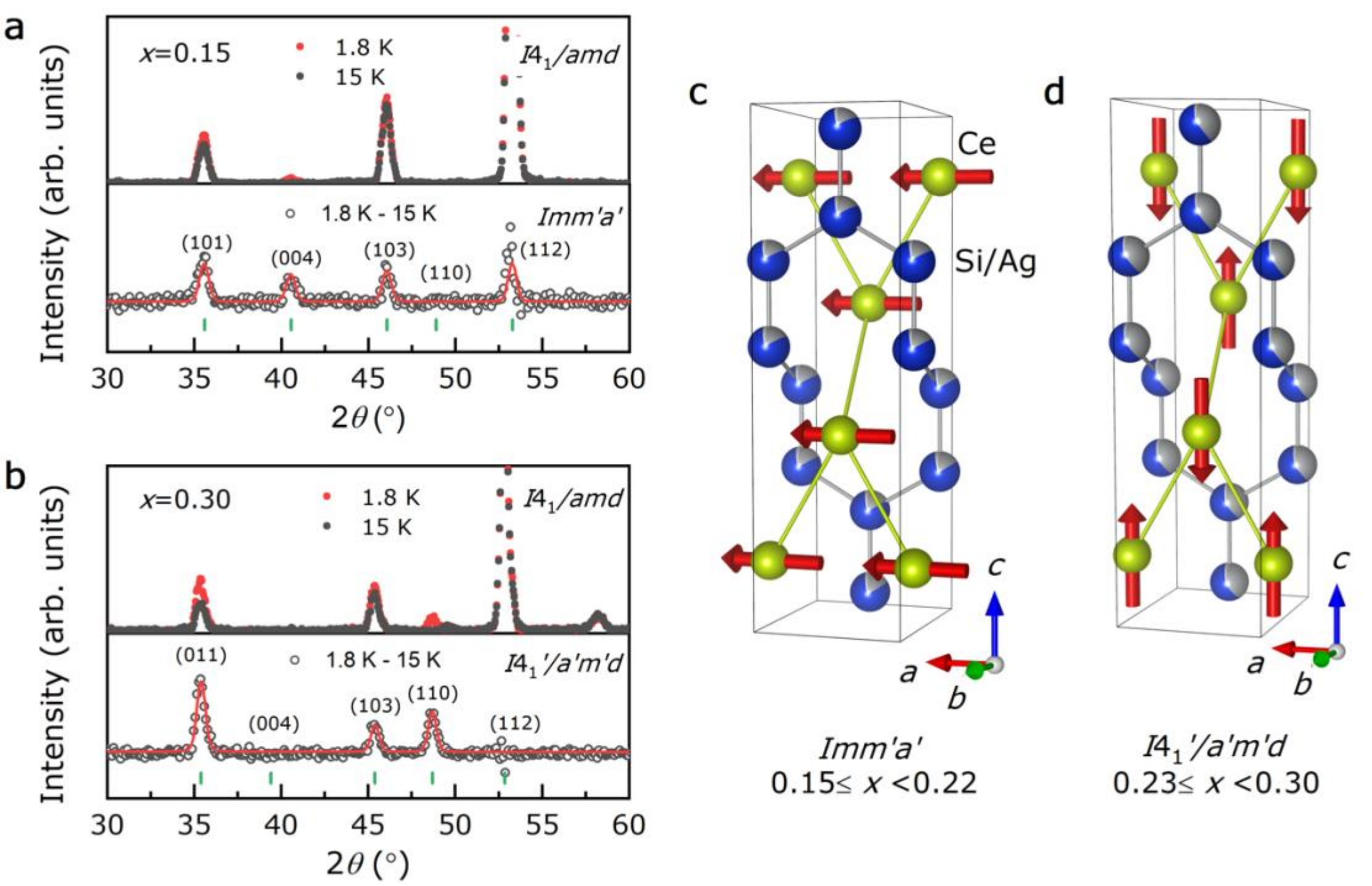

**Extended Data Fig. E2. Neutron powder diffraction results and obtained magnetic structures.** **a (b),** In the upper panel, the black and red colour symbols are neutron powder diffraction patterns of $x$ = 0.15 ($x$ = 0.30) collected at $T$ = 15 and 1.8 K, respectively. All Bragg peaks are assigned to a structure space group $I4_1/amd$. In the lower panel, the black open symbols represent the difference pattern of $x$ = 0.15 ($x$ = 0.30) obtained by subtracting the pattern collected at $T$ = 15 K from the one at $T$ = 1.8 K. The red line in the lower panel is a fitting result using the Shubnikov magnetic space group *Imm'a'* ($I4_1'/a'm'd$) for $x$ = 0.15 ($x$ = 0.30). **c (d)**, Schematic for the magnetic structure was obtained from Rietveld refinement using *Imm'a'* ($I4_1'/a'm'd$) for FM phase 0.15 ≤ x < 0.22 (for AFM phase 0.23 ≤ x ≤ 0.30). Greenish, blue, and grey colour spheres represent Ce, Si, and Ag atoms, respectively. Red colour arrows represent the ordered magnetic moments in the FM (AFM) phase.

**Extended Data Table E1. Refinement results of neutron powder diffraction data for *x* = 0.15, 0.30, and 0.35 of $Ce(Si_{1-x}Ag_x)_{1.9}$.**

| x | SG [a] | MSG [b] | *a* (Å) | *c* (Å) | *m* | $\chi^2$ [c] |
|---|---|---|---|---|---|---|
| 0.15 | $I4_1/amd$ | *Imm'a'* | 4.1962(1) | 14.1808(3) | 0.69(2) | 1.03/1.03 |
| 0.30 | $I4_1/amd$ | $I4_1'/a'm'd$ | 4.2061(1) | 14.4714(3) | 1.11(1) | 1.01/1.01 |
| 0.35 | $I4_1/amd$ | $I4_1'/a'm'd$ | 4.2098(2) | 14.5782(8) | 1.29(1) | 1.11/1.11 |

a Structure space group used for fitting the pattern collected at *T* = 15 K.
b Magnetic space group used for fitting the pattern collected at *T* = 1.8 K.
c The goodness of fit for difference pattern using the Le Bail/Rietveld refinement.

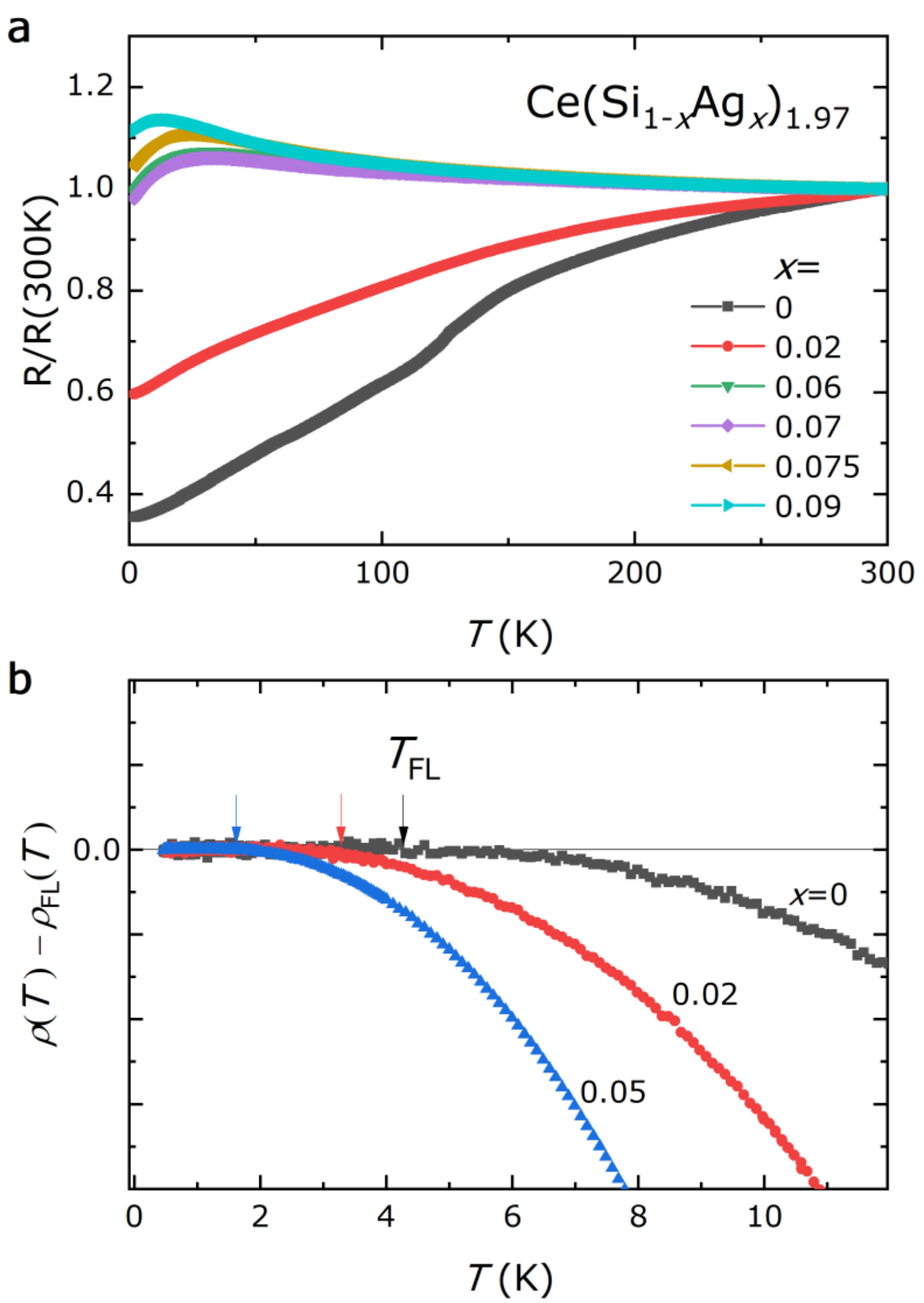

**Extended Data Fig. E3. Normalised electrical resistivity and Fermi liquid behaviour. a,** Electrical resistivity $\rho$ normalised by $\rho(T\text{=300K})$ is plotted as a function of temperature. **b,** Fermi-liquid temperature $T_{\mathrm{FL}}$ below which $\rho(T) = \rho_{\mathrm{FL}}(T) = \rho_0 + \mathrm{A}T^2$ was determined by the temperature where $\rho(T) - \rho_{\mathrm{FL}}(T) = 0$, indicated by downward arrows.

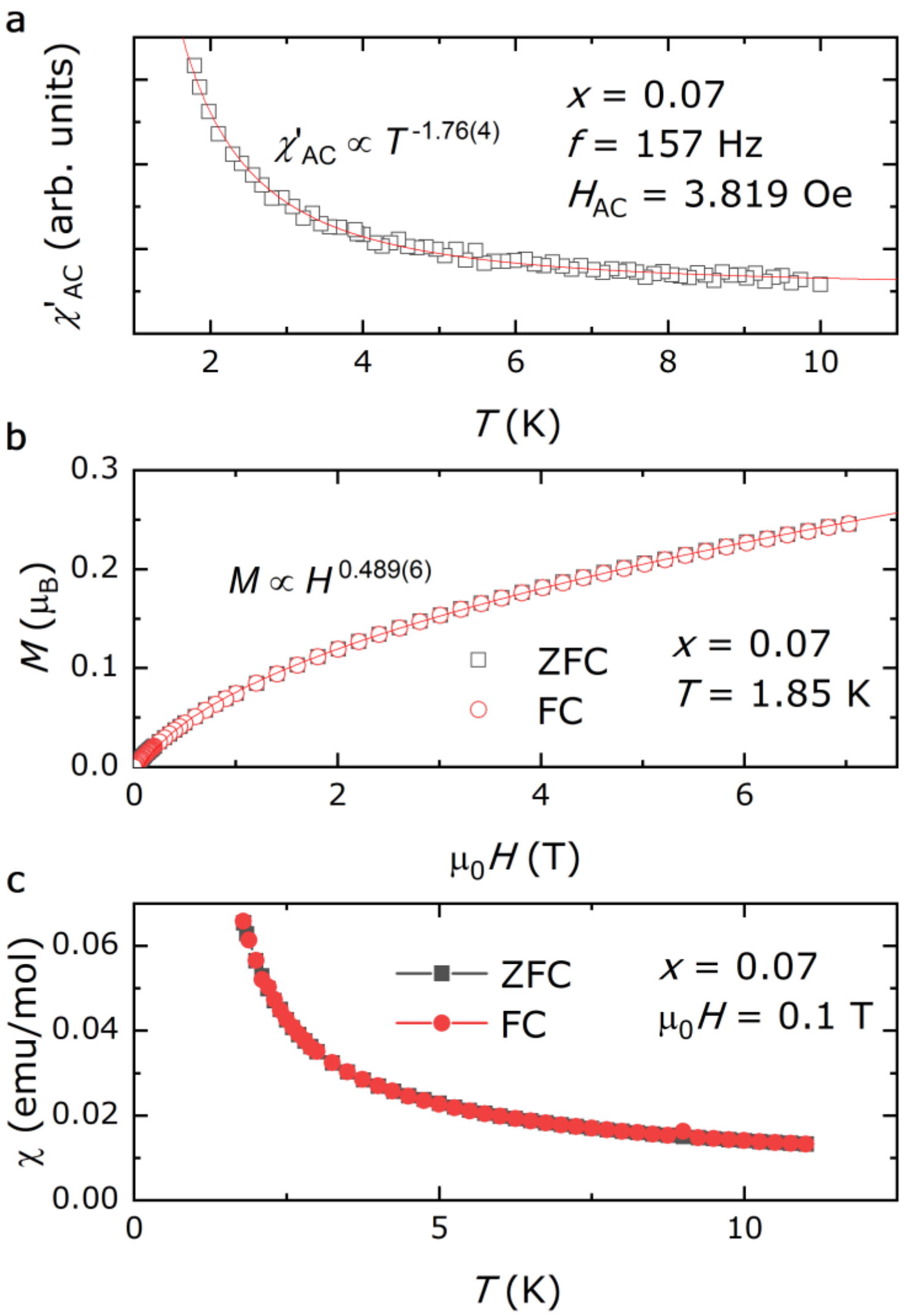

**Extended Data Fig. E4. The least-squares fittings of $\chi'_{AC}(T)$ and $M(H)$, and $\chi(T)$ for $x$ = 0.07. a,** The real part of the AC magnetic susceptibility $\chi'_{AC}(T)$ for $x$ = 0.07 was fitted using the equation of $\chi'_{AC}(T) = a + bT^c$, and the temperature exponent of $c$ = -1.76(4) was obtained. **b,** Field-dependent magnetisation $M(H)$ for $x$ = 0.07 was fitted using $M(H) = d + eH^g$, and the field exponent of $g$ = 0.489(6) was obtained. **c,** The temperature-dependent magnetic susceptibility $\chi(T)$ for $x$ = 0.07 was measured with a zero-field-cooled (ZFC) and field-cooled (FC) process, showing the absence of difference between ZFC and FC data.

**Extended Data Table E2. Single-crystal x-ray diffraction (XRD) results obtained at $T$ = 120 K of $Ce(Si_{1-x}Ag_x)_{2-\delta}$.**

| Nominal $x$ | Refined $x$ | Refined $\delta$ | $a$ (Å) | $c$ (Å) |
|---|---|---|---|---|
| 0 | 0 | 0.034(3) | 4.1660(7) | 13.967(3) |
| 0.02 | 0.020(4) | 0.008(8) | 4.1733(9) | 13.997(5) |
| 0.05 | 0.041(3) | 0.018(6) | 4.1792(8) | 14.065(4) |
| 0.07 | 0.055(1) | 0.045(2) | 4.1904(3) | 14.051(1) |
| 0.10 | 0.091(2) | 0.018(4) | 4.2001(5) | 14.116(1) |
| 0.15 | 0.135(3) | 0.030(6) | 4.2015(4) | 14.192(1) |
| 0.20 | 0.185(2) | 0.030(4) | 4.2092(11) | 14.351(6) |
| 0.25 | 0.227(1) | 0.046(2) | 4.2139(6) | 14.450(3) |
| 0.30 | 0.252(1) | 0.096(2) | 4.2214(4) | 14.471(3) |
| 0.35 | 0.300(2) | 0.100(4) | 4.2244(7) | 14.659(3) |

## Supplementary Information

# Ferromagnetic quantum critical point in a locally non-centrosymmetric and non-symmorphic Kondo metal

Soohyeon Shin,[1,*] Aline Ramires,[2,*] Vladimir Pomjakushin,[3]
Igor Plokhikh,[1] Marisa Medarde,[1] and Ekaterina Pomjakushina[1]

[1]*Laboratory for Multiscale Materials Experiments, Paul Scherrer Institut, Villigen 5232, Switzerland*

[2]*Condensed Matter Theory Group, Paul Scherrer Institut, Villigen 5232, Switzerland*

[3]*Laboratory for Neutron Scattering and Imaging, Paul Scherrer Institut, Villigen 5232 PSI, Switzerland*

## I. THEORETICAL DISCUSSION

$Ce(Si_{1-x}Ag_x)_{1.9}$ crystallizes in the $ThSi_2$-type tetragonal structure with space group $I4_1/amd$ (#141), which is globally centrosymmetric. Note, thought, that none of the atomic sites are centers of inversion, characterizing this system as locally noncentrosymmetric. The metallic state at low temperatures displays a large specific heat coefficient of the order of $0.1J/molK^2$, indicating that this is a heavy Fermi liquid, with effective quasiparticles that carry spectral weight from f-electrons associated with the Ce atoms. This observation motivates us to model the low-temperature metallic state based on effective orbitals at the two inequivalent Ce sublattices.

The minimal Hamiltonian for a locally noncentrosymmetric system can be written in terms of Pauli matrices $\tau_i$ and $\sigma_i$ ($i = 1, 2, 3$), and the corresponding two-dimensional identity matrices $\tau_0$ and $\sigma_0$, encoding the sublattice $(1, 2) = (\mathrm{Ce}, \mathrm{Ce}')$ and spin $(\uparrow, \downarrow)$ degree of freedom (DOF), respectively:

$$H = \sum_{\mathbf{k}} \Psi_{\mathbf{k}}^\dagger [\xi_{\mathbf{k}} \tau_0 \otimes \sigma_0 + \Delta_{1\mathbf{k}} \tau_1 \otimes \sigma_0 + \Delta_{2\mathbf{k}} \tau_2 \otimes \sigma_0 + \mathbf{g}_{\mathbf{k}} \cdot (\tau_3 \otimes \boldsymbol{\sigma})] \Psi_{\mathbf{k}}, \quad (1)$$

with the basis $\Psi_{\mathbf{k}}^\dagger = (c_{\mathbf{k}1\uparrow}^\dagger, c_{\mathbf{k}1\downarrow}^\dagger, c_{\mathbf{k}2\uparrow}^\dagger, c_{\mathbf{k}2\downarrow}^\dagger)$. Here $c_{\mathbf{k}\alpha\beta}^\dagger$ is a creation operator for an electron with momentum $\mathbf{k}$ in sublattice $\alpha$ with spin $\beta$. $\xi_{\mathbf{k}}$ corresponds to the intra-sublattice hopping, $\Delta_{i\mathbf{k}}$ correspond to inter-sublattice hopping (ISH), and $\mathbf{g}_{\mathbf{k}}$ encodes spin-orbit coupling (SOC). As inversion symmetry exchanges sublattices ($P = \tau_1 \otimes \sigma_0$), the Hamiltonian respects global inversion

* These two authors contributed equally

symmetry if $\xi_{\mathbf{k}}$ and $\Delta_{1\mathbf{k}}$ are even and $\Delta_{2\mathbf{k}}$ and $\mathbf{g}_{\mathbf{k}}$ are odd in momentum. Note that the sublattice DOF introduced here is in direct correspondence to the chirality DOF discussed by Kirkpatrick and Belitz [1–3].

Below we give the details of the construction of the lowest order terms based on a tight-binding model for the Ce sites for materials in the family of $CeSi_2$. The crystal structure is body centered tetragonal, with lattice vectors $\mathbf{t}_1 = (-a/2, a/2, c/2)$, $\mathbf{t}_2 = (a/2, -a/2, c/2)$, and $\mathbf{t}_3 = (a/2, a/2, -c/2)$, where $a$ and $c$ are the in-plane and out-of-plane lattice constants, respectively. The nearest neighbours Ce sites in the same sublattice are within the xy-plane, separated by $\boldsymbol{\eta}_1 = (a, 0, 0)$, $\boldsymbol{\eta}_2 = (-a, 0, 0)$, $\boldsymbol{\eta}_3 = (0, a, 0)$, and $\boldsymbol{\eta}_4 = (0, -a, 0)$. The lowest order terms in the tight-binding picture take the form (same form for $H_{Ce'-Ce'}$):

$$
\begin{aligned}
H_{Ce-Ce} &= t \sum_{\langle i,j\rangle,\sigma} c^{\dagger}_{i1\sigma} c_{j1\sigma} + h.c. \\
&= t \sum_{\langle i,j\rangle,\sigma} \sum_{\mathbf{k}} c^{\dagger}_{\mathbf{k}1\sigma} e^{-i\mathbf{k}\cdot\mathbf{r}_i} \sum_{\mathbf{k}'} c_{\mathbf{k}'1\sigma} e^{i\mathbf{k}'\cdot\mathbf{r}_j} + h.c. \\
&= t \sum_{\mathbf{k},\mathbf{k}',\sigma} c^{\dagger}_{\mathbf{k}1\sigma} c_{\mathbf{k}'1\sigma} \sum_i e^{-i(\mathbf{k}-\mathbf{k}')\cdot\mathbf{r}_i} \sum_{n=1}^{4} e^{i\mathbf{k}'\cdot\eta_n} + h.c. \\
&= t \sum_{\mathbf{k},\sigma} c^{\dagger}_{\mathbf{k}1\sigma} c_{\mathbf{k}1\sigma} \left[ e^{ik_x a} + e^{-ik_x a} + e^{ik_y a} + e^{-ik_y a} \right] + h.c. \\
&= \sum_{\mathbf{k},\sigma} c^{\dagger}_{\mathbf{k}1\sigma} c_{\mathbf{k}1\sigma} 2t \left[ \cos(k_x a) + \cos(k_y a) \right] + h.c.,
\end{aligned}
\tag{2}
$$

The last form allows us identify $\xi_{\mathbf{k}} = 2t_{11}\left[\cos(k_x a) + \cos(k_y a)\right] - \mu$, where $t$ is the corresponding hopping amplitude and $\mu$ the chemical potential.

Assuming a Ce site is located at the origin, the nearest Ce$'$ sites are located at $\boldsymbol{\delta}_1 = (a/2, 0, c/4)$, $\boldsymbol{\delta}_2 = (0, a/2, -c/4)$, $\boldsymbol{\delta}_3 = (0, -a/2, -c/4)$, and $\boldsymbol{\delta}_4 = (-a/2, 0, c/4)$. Based on this information, we can write down the ISH from Ce to Ce$'$ atoms:

$$
\begin{aligned}
H_{Ce-Ce'} &= t' \sum_{\langle i,j\rangle,\sigma} c^{\dagger}_{i1\sigma} c_{j2\sigma} + h.c. \\
&= t' \sum_{\langle i,j\rangle,\sigma} \sum_{\mathbf{k}} c^{\dagger}_{\mathbf{k}1\sigma} e^{-i\mathbf{k}\cdot\mathbf{r}_i} \sum_{\mathbf{k}'} c_{\mathbf{k}'2\sigma} e^{i\mathbf{k}'\cdot\mathbf{r}_j} + h.c. \\
&= t' \sum_{\mathbf{k},\mathbf{k}',\sigma} c^{\dagger}_{\mathbf{k}1\sigma} c_{\mathbf{k}'2\sigma} \sum_i e^{-i(\mathbf{k}-\mathbf{k}')\cdot\mathbf{r}_i} \sum_{n=1}^{4} e^{i\mathbf{k}'\cdot\delta_n} + h.c. \\
&= t' \sum_{\mathbf{k},\sigma} c^{\dagger}_{\mathbf{k}1\sigma} c_{\mathbf{k}2\sigma} \left[ e^{i(k_x a/2 + k_z c/4)} + e^{i(k_y a/2 - k_z c/4)} + e^{i(-k_y a/2 - k_z c/4)} + e^{i(-k_x a/2 + k_z c/4)} \right] + h.c. \\
&= \sum_{\mathbf{k},\sigma} c^{\dagger}_{\mathbf{k}1\sigma} c_{\mathbf{k}2\sigma} 2t' \left[ e^{ik_z c/4} \cos(k_x a/2) + e^{-ik_z c/4} \cos(k_y a/2) \right] + h.c.,
\end{aligned}
\tag{3}
$$

where $t'$ is the corresponding hopping amplitude. The last form allows us to identify $\Delta_{1\mathbf{k}} = 2t'\cos(k_zc/4)[\cos(k_xa/2)+\cos(k_ya/2)]$ and $\Delta_{2\mathbf{k}} = 2t'\sin(k_zc/4)[\cos(k_xa/2)-\cos(k_ya/2)]$.

The x- and y-components of the SOC can be derived considering an effective staggered electric field generated by the noncentrosymemtric paths between nearest Ce atoms in the same sublattice. The staggered electric field is generated by the Si atoms and has a component in the z-direction. The electric field changes sign from one Ce sublattice to another, and can be written as:

$$\begin{aligned} H_{SOCx} &= [\delta_{1,m}-\delta_{2,m}] \sum_{\langle i,j\rangle,\sigma,\sigma'} \alpha c^{\dagger}_{im\sigma}[\sigma_x]_{\sigma\sigma'}c_{jm\sigma'} + h.c. \\ &= [\delta_{1,m}-\delta_{2,m}] \sum_{\mathbf{k},\sigma,\sigma'} \alpha c^{\dagger}_{\mathbf{k}m\sigma}[\sigma_x]_{\sigma\sigma'}c_{\mathbf{k}m\sigma'}[e^{ik_ya}-e^{-ik_ya}] + h.c. \\ &= [\delta_{1,m}-\delta_{2,m}] \sum_{\mathbf{k},\sigma,\sigma'} c^{\dagger}_{\mathbf{k}m\sigma}[\sigma_x]_{\sigma\sigma'}c_{\mathbf{k}m\sigma'}2i\alpha\sin(k_ya) + h.c., \end{aligned} \tag{4}$$

and

$$\begin{aligned} H_{SOCy} &= [\delta_{1,m}-\delta_{2,m}] \sum_{\langle i,j\rangle,\sigma,\sigma'} \alpha c^{\dagger}_{im\sigma}[\sigma_y]_{\sigma\sigma'}c_{jm\sigma'} + h.c. \\ &= [\delta_{1,m}-\delta_{2,m}] \sum_{\mathbf{k},\sigma,\sigma'} \alpha c^{\dagger}_{\mathbf{k}m\sigma}[\sigma_y]_{\sigma\sigma'}c_{\mathbf{k}m\sigma'}[e^{-ik_xa}-e^{-ik_xa}] + h.c. \\ &= [\delta_{1,m}-\delta_{2,m}] \sum_{\mathbf{k},\sigma,\sigma'} c^{\dagger}_{\mathbf{k}m\sigma}[\sigma_y]_{\sigma\sigma'}c_{\mathbf{k}m\sigma'}(-2i)\alpha\sin(k_xa) + h.c., \end{aligned} \tag{5}$$

from what we can identify $g_x(\mathbf{k}) = 2\alpha\sin(k_ya)$ and $g_y(\mathbf{k}) = -2\alpha\sin(k_xa)$.

The z-component of the SOC can be derived considering an effective tight binding model with a staggered electric field connecting bonds linking next-next-nearest neighbours in the plane. In this case the effective electric field can be thought of as generated by the complementary Ce sublattice and lies on the xy-plane:

$$\begin{aligned} H_{SOCz} &= [\delta_{1,m}-\delta_{2,m}] \sum_{\langle i,j\rangle',\sigma,\sigma'} \alpha c^{\dagger}_{im\sigma}[\sigma_y]_{\sigma\sigma'}c_{jm\sigma'} + h.c. \\ &= [\delta_{1,m}-\delta_{2,m}] \sum_{\mathbf{k},\sigma,\sigma'} \beta c^{\dagger}_{\mathbf{k}m\sigma}[\sigma_z]_{\sigma\sigma'}c_{\mathbf{k}m\sigma'}[e^{i(2k_x+k_y)a}-e^{i(k_x+2k_y)a}+e^{i(-k_x+2k_y)a}-e^{i(-2k_x+k_y)a} \\ &\quad +e^{i(-2k_x-k_y)a}-e^{i(-k_x-2k_y)a}+e^{i(k_x-2k_y)a}-e^{i(2k_x-k_y)a}] \\ &= [\delta_{1,m}-\delta_{2,m}] \sum_{\mathbf{k},\sigma,\sigma'} c^{\dagger}_{\mathbf{k}m\sigma}[\sigma_z]_{\sigma\sigma'}c_{\mathbf{k}m\sigma'}4\beta[\sin(k_xa)\sin(2k_ya)-\sin(k_ya)\sin(2k_xa)], \end{aligned} \tag{6}$$

from what we can identify $g_z(\mathbf{k}) = 4\beta[\sin(k_xa)\sin(2k_ya)-\sin(k_ya)\sin(2k_xa)]$.

Note that the ISH $\Delta_{1\mathbf{k}}$ and $\Delta_{2\mathbf{k}}$ are both zero at some high symmetry points at the Brillouin zone (BZ) edge:

- At lines on the $k_z = 2\pi/c$ plane with $k_x = \pm k_y$. Note that this includes the high symmetry point $Z = (0, 0, 2\pi/c)$;
- At lines with $(k_x, k_y) = (\pm\pi/a, \pm\pi/a)$ for any $k_z$. Note that this includes the high symmetry points $X = (\pi/a, \pi/a, 0)$ and $P = (\pi/a, \pi/a, \pi/c)$;
- For $k_z = 0$ along a line passing through the $X$ point, characterized by the condition $\cos(k_x a/2) = -\cos(k_y a/2)$. Note that this line is an extension of the diagonal lines on the $k_z = 2\pi/c$ plane if one considers the stacking of the BZ for the body centered tetragonal system.

In summary, the ISH term is strictly zero at the high symmetry points $Z$, $X$, and $P$. It is also zero along the horizontal high symmetry lines connecting the $Z$ and $X$ points and along the vertical high symmetry lines passing through the $X$ and $P$ points.

Note that the SOC components also have zeros, but in different regions of the BZ:

- The x-SOC term, proportional to $\sin(k_y a)$, is zero for $k_y = \{0, \pm\pi/a\}$, $\forall k_x$ and $k_z$.
- The x-SOC term, proportional to $\sin(k_x a)$, is zero for $k_x = \{0, \pm\pi/a\}$, $\forall k_y$ and $k_z$.
- The z-SOC term, proportional to $\sin(k_x a)\sin(2k_y a) - \sin(k_y a)\sin(2k_x a)$, is zero along the planes with $k_x = 0$, $k_y = 0$, and $k_x = \pm k_y$, $\forall k_z$. Is it also zero at the planes with $k_x = \pm\pi/a$ and $k_y = \pm\pi/a$, where the x- and y- components of SOC are zero. Furthermore, the z-SOC term is zero for $k_z = 0$ along a line passing through the $X$ point, and along the vertical line passing through the X and P points, where $H_{Ce-Ce'}$ also vanishes.

Figure S1 summarizes these results, showing in dark color the regions at the BZ boundary where the SOC dominates over ISH.

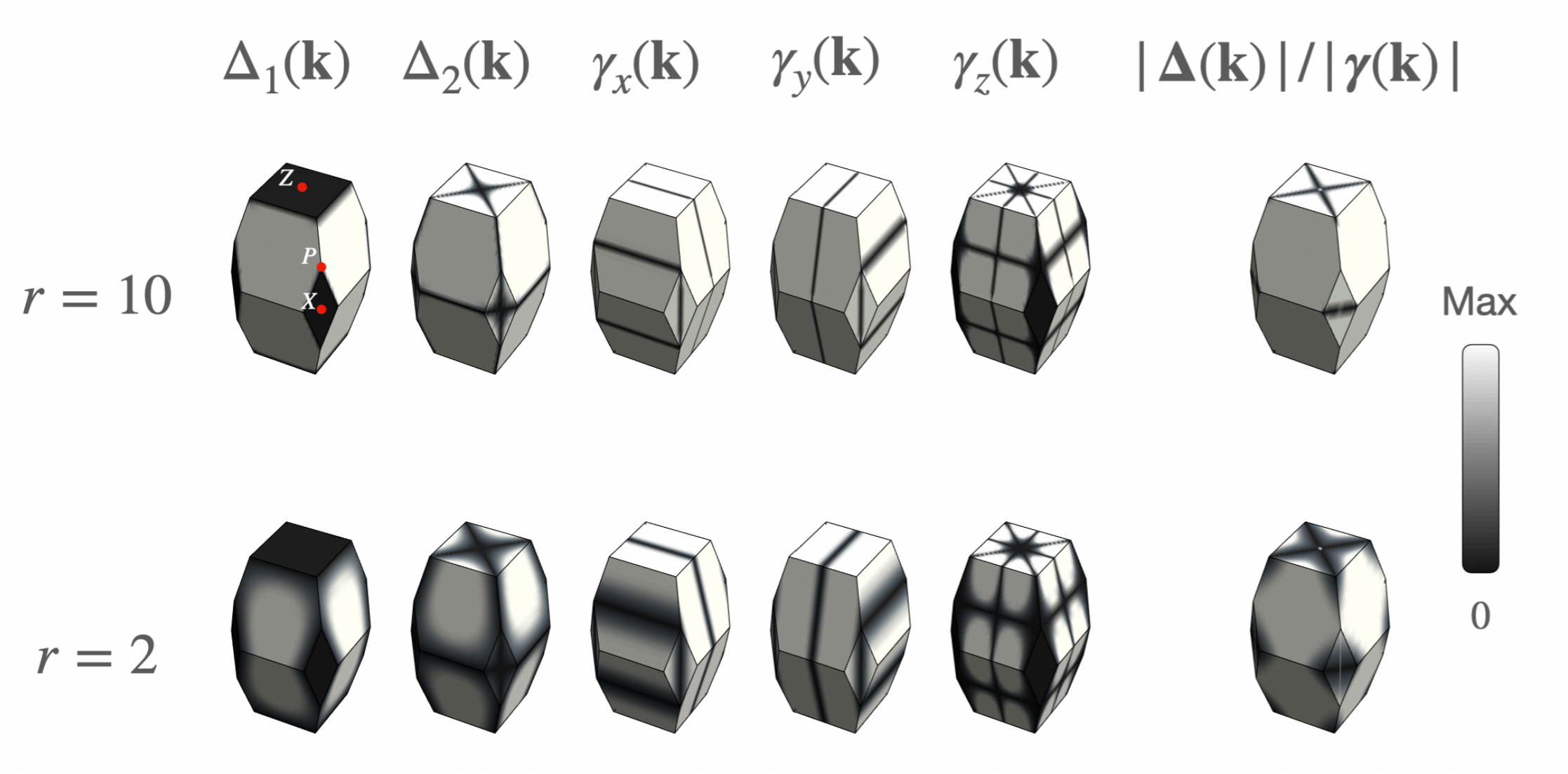

FIG. S1. Brillouin zone (BZ) of $CeSi_2$, a body-centered tetragonal system, and the respective form factors associated with inter-sublattice hopping (ISH) processes, $\boldsymbol{\Delta}(\mathbf{k}) = \{\Delta_1(\mathbf{k}), \Delta_2(\mathbf{k})\}$, and spin-orbit coupling (SOC), $\boldsymbol{\gamma}(\mathbf{k}) = \{\gamma_x(\mathbf{k}), \gamma_y(\mathbf{k}), \gamma_z(\mathbf{k})\}$. The colour scheme is such that white means maximum (up to normalization for each term) and black means zero. The rightmost panels give the form factor of the ratio $|\boldsymbol{\Delta}(\mathbf{k})|/|\boldsymbol{\gamma}(\mathbf{k})|$ for two values of $r = t'/\alpha$ and $\beta = 0$, indicating that ISH can be parametrically smaller than SOC in certain regions at the BZ surfaces. For small values of $r = t'/\alpha$, these regions become more extended. The red points in the top left BZ indicate the high symmetry points.

## II. SUPPLEMENTS FOR EXPERIMENTAL RESULTS

### A. single-crystal x-ray diffraction results

Crystals of $Ce(Si_{1-x}Ag_x)_{2-\delta}$ were mounted on the MiTeGen MicroMounts loop and used for x-ray structure determination. Measurements were performed at $T = 120$ K using the STOE STADIVARI diffractometer equipped with a Dectris EIGER 1M 2R CdTe detector and with an Anton Paar Primux 50 Ag/Mo dual-source using Mo $K_\alpha$ radiation ($\lambda = 0.71073$ Å) from a micro-focus x-ray source and coupled with an Oxford Instruments Cryostream 800 jet.

Data reduction was performed with $X-Area$ package (Version 2.1, STOE & Cie GmbH, Darmstadt, Germany, 2022). The intensities were corrected for Lorentz and polarization effects, and frame scaling along with an empirical absorption correction using spherical harmonics was applied using $X-Area$ package. Crystal structures were solved using charge flipping algorithm implemented in Superflip [4] supplied with JANA2020 [5].

**Table S1. Details of single-crystal diffraction experiments for $Ce(Si_{1-x}Ag_x)_{2-\delta}$. Measurements done at T = 120 K, measured with $MoK_\alpha$ radiation (0.71073Å), space group *I*4₁/*amd* (tetragonal, No. 141, origin choice 1).**

| | x = 0 | x = 0.02 | x = 0.05 | x = 0.07 | x = 0.10 |
|---|---|---|---|---|---|
| *a* in Å | 4.1660(7) | 4.1733(9) | 4.1792(8) | 4.1904(3) | 4.2001(5) |
| *c* in Å | 13.967(3) | 13.997(5) | 14.065(4) | 14.0517(13) | 14.1163(17) |
| *V* in $Å^3$ | 242.41(7) | 243.77(11) | 245.66(10) | 246.74(3) | 249.02(5) |
| $R_{int}$ | 0.0113 | 0.0268 | 0.022 | 0.0189 | 0.0219 |
| $\Delta_{max}/\Delta_{min}$ | 0.47e/<br>-0.41e | 1.18e/<br>-1.25e | 1.26e/<br>-1.71e | 1.26e/<br>-1.27e | 2.36e/<br>-2.35e |
| h, k, l range | -7 < h < 7<br>-5 < k < 7<br>-23 < l < 23 | -7 < h < 7<br>-6 < k < 4<br>-23 < l < 23 | -6 < h < 7<br>-6 < k < 7<br>-22 < l < 23 | -7 < h < 6<br>-6 < k < 7<br>-24 < l < 23 | -6 < h < 7<br>-7 < k < 6<br>-23 < l < 23 |
| θ range | 5.11 - 37.41 | 5.1 - 37.18 | 5.09 - 37.18 | 5.08 - 37.72 | 5.06 - 37.26 |
| N reflections all/<br>merged/<br>> 3σ | 3920/<br>192/<br>158 | 5520/<br>192/<br>158 | 5544/<br>193/<br>159 | 7105/<br>201/<br>165 | 6414/<br>195/<br>163 |
| D in g ·$cm^3$ /<br>μ in $mm^{-1}$ | 5.3532/<br>19.323 | 5.4355/<br>19.523 | 5.4864/<br>19.685 | 5.5225/<br>19.801 | 5.6294/<br>20.148 |
| $R_F$( >3σ)/<br>$wR_F$(> 3σ)/<br>$R_F$(all)/<br>$wR_F$(all) | 0.0092/<br>0.0214/<br>0.0181/<br>0.0227 | 0.0133/<br>0.0319/<br>0.0209/<br>0.0334 | 0.0116/<br>0.0258/<br>0.0211/<br>0.0269 | 0.0045/<br>0.0084/<br>0.0134/<br>0.0090 | 0.0065/<br>0.0152/<br>0.0172/<br>0.0175 |
| $\chi^2(>3\sigma)/\chi^2(all)$ | 1.0/1.04 | 1.0/1.06 | 1.05/1.11 | 1.01/1.05 | 1.07/1.02 |

| | x = 0.15 | x = 0.20 | x = 0.25 | x = 0.30 | x = 0.35 |
|---|---|---|---|---|---|
| *a* in Å | 4.2015(4) | 4.2092(11) | 4.2139(6) | 4.2214(4) | 4.2244(7) |
| *c* in Å | 14.1920(19) | 14.351(6) | 14.450(3) | 14.471(2) | 14.659(3) |
| *V* in $Å^3$ | 250.53(5) | 254.26(14) | 256.58(7) | 257.88(5) | 261.60(9) |
| $R_{int}$ | 0.0272 | 0.0148 | 0.0274 | 0.0257 | 0.0344 |
| $\Delta_{max}/\Delta_{min}$ | 1.19e/<br>-1.15e | 0.56e/<br>-0.68e | 0.67e/<br>-0.65e | 0.83e/<br>-0.81e | 1.24e/<br>-0.69e |
| h, k, l range | -6 < h < 7<br>-7 < k < 5<br>-24 < l < 24 | -7 < h < 5<br>-7 < k < 7<br>-24 < l < 23 | -7 < h < 7<br>-7 < k < 7<br>-17 < l < 23 | -7 < h < 7<br>-4 < k < 7<br>-24 < l < 24 | -6 < h < 6<br>-6 < k < 6<br>-19 < l < 23 |
| θ range | | 5.05 - 37.35 | 5.04 - 37.15 | 5.03 - 37.48 | 5.02 - 34.77 |
| N reflections<br>all/merged/above<br>3σ | 5751/<br>196/<br>162 | 3957/<br>204/<br>166 | 5626/<br>200/<br>163 | 3891/<br>206/<br>169 | 4243/<br>178/<br>142 |
| D in g ·$cm^{-3}$ /<br>μ in $mm^{-1}$ | 5.7754/<br>20.631 | 5.9027/<br>21.04 | 6.0192/<br>21.421 | 6.0936/<br>21.665 | 6.1993/<br>22.003 |
| $R_F$( >3σ)/<br>$wR_F$(> 3σ)/<br>$R_F$(all)/<br>$wR_F$(all) | 0.0126/<br>0.0345/<br>0.0204/<br>0.0371 | 0.0084/<br>0.0136/<br>0.0137/<br>0.0136 | 0.0068/<br>0.0106/<br>0.0145/<br>0.0107 | 0.0086<br>0.0189<br>0.0155<br>0.0193 | 0.0135/<br>0.0301/<br>0.0203/<br>0.0307 |
| $\chi^2(>3\sigma)/\chi^2(all)$ | 1.01/1.04 | 0.99/1.1 | 1.03/1.13 | 1.0/1.09 | 0.99/1.1 |

**Table S2. Atomic coordinates and displacement parameters. Ce is in 4e (0,0,0) position, Si/Ag mixed site 8e (0, 0, z).**

| Atom | Occupancy | x/a | y/b | z/c | $U_{iso}$ |
|---|---|---|---|---|---|
| | | x = 0 | | | |
| Ce1 | 1 | 0 | 0 | 0 | 0.00406(5) |
| Si1 | 0.983(6) | 0 | 0 | 0.58399(6) | 0.0069(2) |
| | | x = 2 | | | |
| Ce1 | 1 | 0 | 0 | 0 | 0.00876(7) |
| Si1 | 0.980(4) | 0 | 0 | 0.58397(7) | 0.0109(3) |
| Ag1 | 0.020(4) | | | | |
| | | x = 5 | | | |
| Ce1 | 1 | 0 | 0 | 0 | 0.00691(6) |
| Si1 | 0.959(3) | 0 | 0 | 0.58393(6) | 0.0088(2) |
| Ag1 | 0.041(3) | | | | |
| | | x = 0.07 | | | |
| Ce1 | 1 | 0 | 0 | 0 | 0.00443(2) |
| Si1 | 0.9447(8) | 0 | 0 | 0.583927(19) | 0.00526(9) |
| Ag1 | 0.0553(8) | | | | |
| | | x =0.10 | | | |
| Ce1 | 1 | 0 | 0 | 0 | 0.00462(4) |
| Si1 | 0.9083(16) | 0 | 0 | 0.58389(3) | 0.00505(13) |
| Ag1 | 0.0917(16) | | | | |
| | | x = 0.15 | | | |
| Ce1 | 1 | 0 | 0 | 0 | 0.00707(8) |
| Si1 | 0.865(3) | 0 | 0 | 0.58367(5) | 0.0081(2) |
| Ag1 | 0.135(3) | | | | |
| | | x = 0.20 | | | |
| Ce1 | 1 | 0 | 0 | 0 | 0.00443(4) |
| Si1 | 0.8141(11) | 0 | 0 | 0.58341(3) | 0.00484(14) |
| Ag1 | 0.1859(11) | | | | |
| | | x =0.25 | | | |
| Ce1 | 1 | 0 | 0 | 0 | 0.00526(4) |
| Si1 | 0.7730(8) | 0 | 0 | 0.583155(19) | 0.00613(11) |
| Ag1 | 0.2270(8) | | | | |
| | | x = 0.30 | | | |
| Ce1 | 1 | 0 | 0 | 0 | 0.00442(5) |
| Si1 | 0.7476(13) | 0 | 0 | 0.58302(3) | 0.00564(13) |
| Ag1 | 0.2524(13) | | | | |
| | | x = 0.35 | | | |
| Ce1 | 1 | 0 | 0 | 0 | 0.00705(9) |
| Si1 | 0.700(2) | 0 | 0 | 0.58310(4) | 0.0083(2) |
| Ag1 | 0.300(2) | | | | |

**Table S3. Components of anisotropic tensor ($U_{12} = U_{23} = U_{13} = 0$).**

| Atom | $U_{11}$ | $U_{22}$ | $U_{33}$ |
|---|---|---|---|
| | x = 0 | | |
| Ce1 | 0.00344(8) | 0.00344(8) | 0.00531(10) |
| Si1 | 0.0056(4) | 0.0072(4) | 0.0078(4) |
| | x = 0.02 | | |
| Ce1 | 0.00734(12) | 0.00734(12) | 0.01159(14) |
| Si1/Ag1 | 0.0094(6) | 0.0095(6) | 0.0137(5) |
| | x = 0.05 | | |
| Ce1 | 0.00584(10) | 0.00584(10) | 0.00905(12) |
| Si1/Ag1 | 0.0083(4) | 0.0075(4) | 0.0107(4) |
| | x = 0.07 | | |
| Ce1 | 0.00419(4) | 0.00419(4) | 0.00490(4) |
| Si1/Ag1 | 0.00591(16) | 0.00485(16) | 0.00502(13) |
| | x = 0.10 | | |
| Ce1 | 0.00422(6) | 0.00422(6) | 0.00542(7) |
| Si1/Ag1 | 0.0062(2) | 0.0044(2) | 0.0046(2) |
| | x = 0.15 | | |
| Ce1 | 0.00611(13) | 0.00611(13) | 0.00900(15) |
| Si1/Ag1 | 0.0092(4) | 0.0067(4) | 0.0084(4) |
| | x = 0.20 | | |
| Ce1 | 0.00381(7) | 0.00381(7) | 0.00568(9) |
| Si1/Ag1 | 0.0035(3) | 0.0070(3) | 0.00401(19) |
| | x = 0.25 | | |
| Ce1 | 0.00471(6) | 0.00471(6) | 0.00634(9) |
| Si1/Ag1 | 0.0086(2) | 0.00457(19) | 0.00516(18) |
| | x = 0.30 | | |
| Ce1 | 0.00458(8) | 0.00458(8) | 0.00410(9) |
| Si1/Ag1 | 0.0081(2) | 0.0050(2) | 0.00382(19) |
| | x = 0.35 | | |
| Ce1 | 0.00628(13) | 0.00628(13) | 0.00861(18) |
| Si1/Ag1 | 0.0107(4) | 0.0061(4) | 0.0080(3) |

### B. Neutron powder diffraction analysis

Here, we address the details of Rietveld refinements for neutron diffraction data. Neutron powder diffraction (NPD) patterns, obtained at $T$ = 15 K above ordering temperatures, were analysed by Rietveld refinement using the FullProf suite [6], and all cases were fit by the room temperature structure $I4_1/amd$ with decreased lattice parameters. As shown in Extended Data Figs. E2a and b, NPD patterns exhibit additional intensities due to the magnetic scattering below ordering temperatures. Difference patterns, obtained by subtracting the pattern collected at $T$ = 15 K from the pattern at $T$ = 1.8 K, were fit by Le Bail model in order to check ordering wave vectors **k**'s. For both ferromagnetic (FM) and antiferromagnetic (AFM) cases, $\mathbf{k} = 0$ shows the best result. Using crystallographic information obtained by Rietveld refinement, possible magnetic space groups were investigated using Bilbao crystallographic server [7] and ISODISTORT tool based on ISOTROPY software [8, 9]. Figure S2 shows all possible magnetic subgroups that give non-zero magnetic moments for magnetic Ce ions located on Wckyoff position $4a$(0, 3/4, 1/8) of spacegroup $I4_1/amd$ (no. 141) when $\mathbf{k} = 0$. As shown in Fig. S2b, six maximal symmetry subgroups were adopted for fitting difference patterns, and the orthorhombic $Imm'a'$ (no. 74.559) and the tetragonal $I4_1'/a'm'd$ (no. 141.556) gives the same fitting quality with Le Bail fittings for ferromagnetic (FM) and antiferromagnetic (AFM) structures, respectively. The obtained magnetic structures are depicted in Extended Data Figs. E2c and d. The unit cell transformations from parent paramagnetic space groups are given by the following relations (A, B, C) = 1/4 + (-b, a, c) in $Imm'a'$ and (A, B, C) = (a, b, c) in $I4_1'/a'm'd$, where the capital letter and lower case are the basis vectors of the magnetic and the parent paramagnetic space group, respectively.

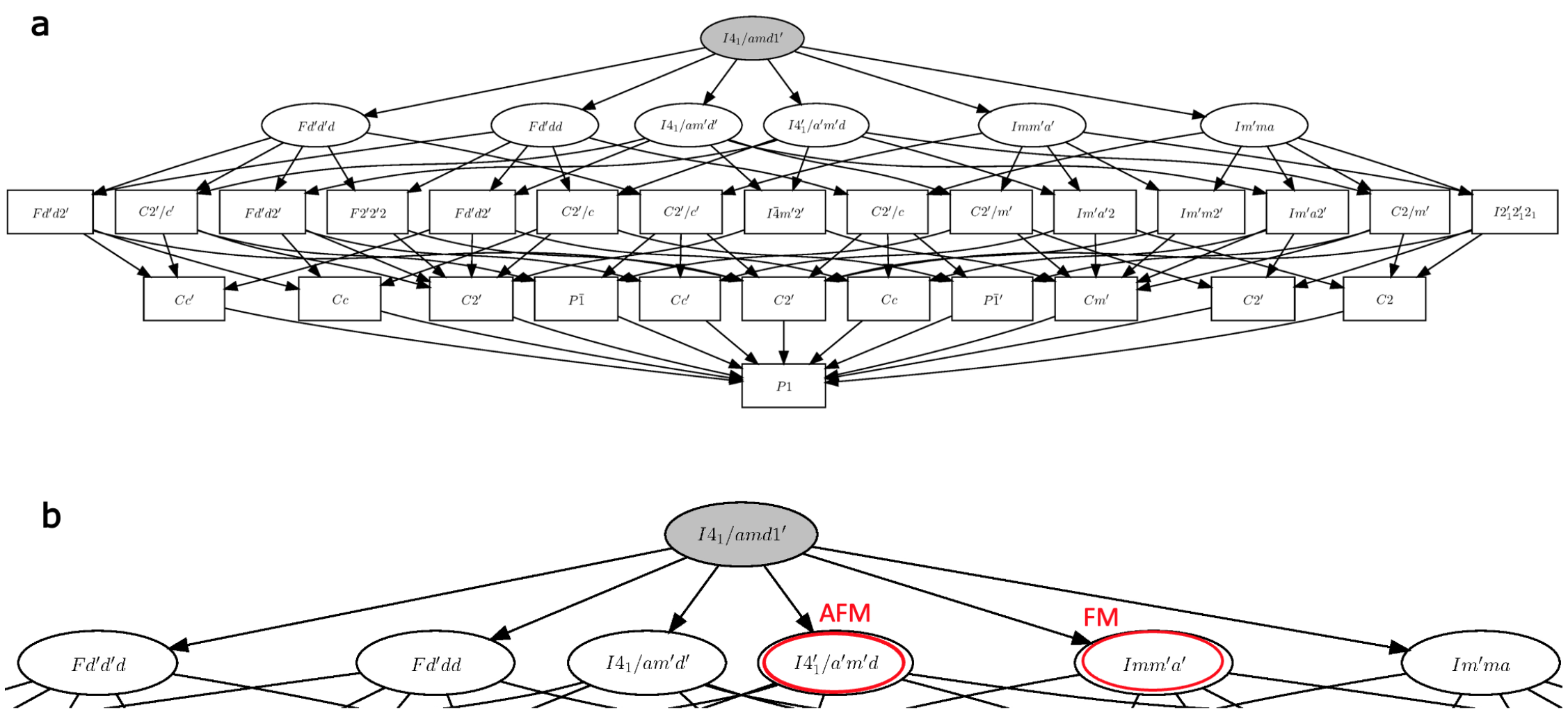

FIG. S2. **Magnetic subgroup graph for $\mathbf{CeSi_{2-d}}$. a,** The graph displays all magnetic subgroups that give non-zero magnetic moments for Ce sites $4a(0,\ 3/4,\ 1/8)$ in a space group $I4_1/amd$ (no. 141) when the magnetic ordering wave vector $\mathbf{k}$ is 0. **b.** Only the maximal magnetic subgroups are displayed and red circles indicate subgroups $Imm'a'$ (no. 74.559) and $I4'_1/a'm'd$ (no. 141.556) giving the best refinement results for ferromagnetic (FM) and antiferromagnetic (AFM) structures, respectively.